\documentclass{amsart}
\usepackage{amsmath,amsthm,amsfonts,amscd,amssymb}
\usepackage{verbatim}
\usepackage[pdfborder={0 0 0}]{hyperref}
\usepackage{color}

\newtheorem{thm}{Theorem}[section]

\newtheorem{lem}[thm]{Lemma}

\theoremstyle{definition}

\theoremstyle{remark}

\begin{document}

\title[One-coincidence sequences for frequency hopping]{A NEW FAMILY OF ONE-COINCIDENCE SETS OF SEQUENCES WITH DISPERSED ELEMENTS FOR FREQUENCY HOPPING CDMA SYSTEMS}

\author{Lenny Fukshansky}
\author{Ahmad A. Shaar}

\address{Department of Mathematics, 850 Columbia Avenue, Claremont McKenna College, Claremont, CA 91711}
\email{lenny@cmc.edu}
\address{Department of Engineering, Harvey Mudd College, Claremont, CA 91711}
\email{ashaar@g.hmc.edu}


\begin{abstract}
We present a new family of one-coincidence sequence sets suitable for frequency hopping code division multiple access (FH-CDMA) systems with dispersed (low density) sequence elements. These sets are derived from one-coincidence prime sequence sets, such that for each one-coincidence prime sequence set there is a new one-coincidence set comprised of sequences with dispersed sequence elements, required in some circumstances, for FH-CDMA systems. Getting rid of crowdedness of sequence elements is achieved by doubling the size of the sequence element alphabet. In addition, this doubling process eases control over the distance between adjacent sequence elements. Properties of the new sets are discussed.
\end{abstract}

\maketitle
%\tableofcontents

\def\A{{\mathcal A}}
\def\AA{{\mathfrak A}}
\def\B{{\mathcal B}}
\def\C{{\mathcal C}}
\def\D{{\mathcal D}}
\def\E{{\mathcal E}}
\def\F{{\mathcal F}}
\def\Ff{{\mathfrak F}}
\def\G{{\mathcal G}}
\def\x{{\mathcal H}}
\def\I{{\mathcal I}}
\def\J{{\mathcal J}}
\def\K{{\mathcal K}}
\def\kk{{\mathfrak K}}
\def\L{{\mathcal L}}
\def\LL{{\mathfrak L}}
\def\M{{\mathcal M}}
\def\O{{\mathcal O}}
\def\W{{\omega}}
\def\CC{{\mathfrak C}}
\def\mm{{\mathfrak m}}
\def\MM{{\mathfrak M}}
\def\OO{{\mathfrak O}}
\def\P{{\mathcal P}}
\def\R{{\mathcal R}}
\def\s{{\mathcal S}}
\def\V{{\mathcal V}}
\def\X{{\mathcal X}}
\def\XX{{\mathfrak X}}
\def\Y{{\mathcal Y}}
\def\Z{{\mathcal Z}}
\def\H{{\mathcal H}}
\def\cee{{\mathbb C}}
\def\pee{{\mathbb P}}
\def\que{{\mathbb Q}}
\def\real{{\mathbb R}}
\def\zed{{\mathbb Z}}
\def\hyp{{\mathbb H}}
\def\aaa{{\mathbb A}}
\def\Nn{{\mathbb N}}
\def\ff{{\mathbb F}}
\def\kk{{\mathfrak K}}
\def\qbar{{\overline{\mathbb Q}}}
\def\kbar{{\overline{K}}}
\def\ybar{{\overline{Y}}}
\def\kkbar{{\overline{\mathfrak K}}}
\def\ubar{{\overline{U}}}
\def\eps{{\varepsilon}}
\def\ahat{{\hat \alpha}}
\def\bhat{{\hat \beta}}
\def\gt{{\tilde \gamma}}
\def\h{{\tfrac12}}
\def\dd{{\partial}}
\def\baa{{\boldsymbol \alpha}}
\def\bfa{{\boldsymbol a}}
\def\bfb{{\boldsymbol b}}
\def\be{{\boldsymbol e}}
\def\bei{{\boldsymbol e_i}}
\def\bff{{\boldsymbol f}}
\def\bc{{\boldsymbol c}}
\def\bm{{\boldsymbol m}}
\def\bk{{\boldsymbol k}}
\def\bi{{\boldsymbol i}}
\def\bj{{\boldsymbol j}}
\def\bl{{\boldsymbol l}}
\def\bq{{\boldsymbol q}}
\def\bu{{\boldsymbol u}}
\def\bt{{\boldsymbol t}}
\def\bs{{\boldsymbol s}}
\def\bfu{{\boldsymbol u}}
\def\bv{{\boldsymbol v}}
\def\bw{{\boldsymbol w}}
\def\bx{{\boldsymbol x}}
\def\bX{{\boldsymbol X}}
\def\bz{{\boldsymbol z}}
\def\bwy{{\boldsymbol y}}
\def\bY{{\boldsymbol Y}}
\def\bL{{\boldsymbol L}}
\def\ba{{\boldsymbol a}}
\def\bb{{\boldsymbol b}}
\def\bet{{\boldsymbol\eta}}
\def\bxi{{\boldsymbol\xi}}
\def\bo{{\boldsymbol 0}}
\def\bol{{\boldkey 1}_L}
\def\ep{\varepsilon}
\def\p{\boldsymbol\varphi}
\def\q{\boldsymbol\psi}
\def\rank{\operatorname{rank}}
\def\aut{\operatorname{Aut}}
\def\lcm{\operatorname{lcm}}
\def\sgn{\operatorname{sgn}}
\def\spn{\operatorname{span}}
\def\md{\operatorname{mod}}
\def\Norm{\operatorname{Norm}}
\def\dim{\operatorname{dim}}
\def\det{\operatorname{det}}
\def\Vol{\operatorname{Vol}}
\def\rk{\operatorname{rk}}
\def\ord{\operatorname{ord}}
\def\ker{\operatorname{ker}}
\def\div{\operatorname{div}}
\def\Gal{\operatorname{Gal}}
\def\GL{\operatorname{GL}}
\def\p{\operatorname{p}}
\def\q{\operatorname{q}}
\def\t{\operatorname{t}}
\def\hs{{\hat \sigma}}
\def\chr{\operatorname{char}}
\def\req{\operatorname{req}}
\def\H{\operatorname{H}}
\def\mp{\operatorname{\cdot_p}}

\section{Introduction}
\label{intro}

In the early 1980s a survey of one-coincidence multilevel sequence sets for FH-CDMA had been presented in~\cite{1}. Each multilevel sequence is used to specify which frequency will be used, by any pair of users (or a single radar), for transmission (reception) at any given time. At the same time, this hopping sequence should be designed in a way that insures minimum mutual interference between different users of the channel. This interference can be measured by the Hamming cross-correlation between different sequences used by different pair of users. Let $\{ F_1,\dots,F_q \}$ be a set of frequencies. For $L \leq q$, let $X = (X_0,\dots,X_{L-1})$ and $Y = (Y_0,\dots,Y_{L-1})$ be two hopping frequency sequences with $X_i,Y_i \in \{ F_1,\dots,F_q \}$ for all $0 \leq i \leq L-1$. The periodic Hamming cross-correlation between the pair of sequences $X$ and $Y$ is defined as in~\cite{3}:
\begin{equation}
\label{HXY}
\H_{XY}(\tau) = \sum_{i=0}^{L-1} h(X_i,Y_{i+\tau}),
\end{equation}
for all $0 \leq \tau < L$, where the subcript $i+\tau$ in the sum above is taken modulo $L$ and
$$h(a,b) = \left\{ \begin{array}{ll}
0 & \mbox{if } a \neq b \\
1 & \mbox{if } a = b \\
\end{array}
\right.$$
Equation \eqref{HXY} represents the number of coincidences (hits) between sequences $X$ and $Y$ for relative time delay $\tau$.

A one-coincidence sequence set is a set of non-repeating elements for which the peak of the Hamming cross-correlation function equals one for any pair of sequences belonging to the set, i.e. the maximum number of hits between any pair of sequences from the set is one for any relative shift~$\tau$. The one-coincidence sequences discussed and constructed in~\cite{1} possess the following properties:

\begin{enumerate}
\item All of the sequences are of the same length.
\item Length $L$ of each sequence is close to the size $q$ of the alphabet used: extremely crowded sequences.
\item All of the sequences are non-repeating, that is, each frequency is used at most once within the sequence period. This property facilitates a simple synchronization scheme.
\item The maximum number of hits between any pair of sequences for any time shift $\tau$ equals one.
\end{enumerate}

Using crowded sequences presents a certain advantage, since in that case the system needs a smaller band of frequencies, providing frequency space for other friendly systems. On the other hand, there is also a disadvantage, because it means heavy mutual interference between friendly systems. Doubling the size of the alphabet gives the chance of less crowded environments  and allows for larger distances between hopped frequencies. To quote~\cite{4}:

{\it ``In the frequency-hopping spread spectrum systems, it is desirable for the transmitter to hop to a frequency far from the previous one, that is, any adjacent frequencies in any sequences are spaced far apart."}

This property can improve the receiver's resistance to various interferences. Therefore, one-coincidence sequences with the property that the distance between any adjacent symbols in any sequence is greater than a specified amount are considered in~\cite{4}. The construction presented in~\cite{4} has a small drawback: it is easy to discover all sequences of the other pairs of users from only one sequence. This could jeopardize the main objective of frequency hopping systems.

Bluetooth systems use adaptive frequency-hopping (AFH) spread spectrum to improve resistance to radio frequency interference by avoiding crowded frequencies in the hopping sequence~\cite{5}. This can be carried out through assigning hopping patterns to users, which have no fading frequencies. This assigning process is much easier by selecting sequences from our non-crowded sequence sets of large size.

This note is organized as follows. In Section~\ref{HMC} we present the construction of a new one-coincidence sequence sets with elements distributed in a non-crowded manner on the axis of positive integers, as derived from prime sequence sets~\cite{1}, such that for each prime sequence set there is a new one-coincidence set comprised of sequences with dispersed element distribution, suitable for frequency hopping systems. This construction is clarified through examples. In Section~\ref{lemmas} we discuss the properties of these sequence sets. We finish in Section~\ref{close} with a short summary of the main results and a question for further research.

One novel feature of our construction is that it combines binary operations on $\zed$ with $\md p$ binary operations on~$\ff_p$. To the best of our knowledge, such a combination has not previously been used in the design of digital sequences. The proofs of the properties of our sequence sets, detailed in Section~\ref{lemmas} explore the interplay between these operations. We believe that this idea may have further applications in the analysis of digital sequences with interesting properties.

\bigskip

\section{HMC sequence set construction and examples}
\label{HMC}

We construct here a new family of sequence sets, which we call HMC after the Harvey Mudd College, acknowledging the support it provided to the second author. Let $p \geq 3$ be a prime and $\ff_p$ be the corresponding prime field, equal to $\{ 0,1, \dots, p-1 \}$ as a set with addition $+_p$ and multiplication $\cdot_p$ operations modulo $p$; in particular, we will distinguish between $a+b$ (usual sum) and $a +_p b$ (sum $\md p$), as well as $ab$ (usual product) and $a \cdot_p b$ (product $\md p$). Then for each $k \in \ff_p$, define a prime sequence
$$S_k = \left\{ 0, k, 2 \cdot_p k,\dots,(p-1) \cdot_p k \right\},$$
which is just a permutation of a fixed ordering of $\ff_p$, given by $\md p$-multiplication by $k$. Now we define the HMC sequence
$$H_k = \left\{ 0+k, k+2 \cdot_p k, 2 \cdot_p k+3 \cdot_p k,\dots,(p-1) \cdot_p k+0 \right\},$$
where multiplication is again $\md p$, but addition is not; in other words, the $j$-th element of the sequence $H_k$ is the sum of $j$-th and $(j+1)$-st elements of $S_k$ viewed as integers for every $1 \leq j \leq p-1$, while the $p$-th element of $H_k$ remains $(p-1) \cdot_p k$. We present examples of this simple construction when $p=7$ in Table~\ref{S-7} (the sequence set $S_0,\dots,S_6$) and Table~\ref{H-7} (the sequence set $H_1,\dots,H_6$).

\begin{center} 
\begin{table}[!ht]
\caption{Set of prime sequences for $p = 7$}
\begin{tabular}{|l||l|l|l|l|l|l|l|} \hline
$j \in \ff_7$ & 0 & 1 & 2 & 3 & 4 & 5 & 6 \\ \hline
Sequence $S_0=0 \cdot_7 j$ & 0 & 0 & 0 & 0 & 0 & 0 & 0 \\ \hline
Sequence $S_1=1\cdot_7 j$ & 0 & 1 & 2 & 3 & 4 & 5 & 6 \\ \hline
Sequence $S_2=2\cdot_7 j$ & 0 & 2 & 4 & 6 & 1 & 3 & 5 \\ \hline
Sequence $S_3=3\cdot_7 j$ & 0 & 3 & 6 & 2 & 5 & 1 & 4 \\ \hline
Sequence $S_4=4\cdot_7 j$ & 0 & 4 & 1 & 5 & 2 & 6 & 3 \\ \hline
Sequence $S_5=5\cdot_7 j$ & 0 & 5 & 3 & 1 & 6 & 4 & 2 \\ \hline
Sequence $S_6=6\cdot_7 j$ & 0 & 6 & 5 & 4 & 3 & 2 & 1 \\ \hline
\end{tabular}
\label{S-7}
\end{table}
\end{center}

\begin{center} 
\begin{table}[!ht]
\caption{Set of HMC sequences for $p = 7$ with minimum distance $d$ between consecutive elements}
\begin{tabular}{|l||l|l|l|l|l|l|l||l|} \hline
Sequence $H_1$ & 1 & 3 & 5 & 7 & 9 & 11 & 6 & $d=2$ \\ \hline
Sequence $H_2$ & 2 & 6 & 10 & 7 & 4 & 8 & 5 & $d=3$\\ \hline
Sequence $H_3$ & 3 & 9 & 8 & 7 & 6 & 5 & 4 & $d=1$\\ \hline
Sequence $H_4$ & 4 & 5 & 6 & 7 & 8 & 9 & 3 & $d=1$\\ \hline
Sequence $H_5$ & 5 & 8 & 4 & 7 & 10 & 6 & 2 & $d=3$\\ \hline
Sequence $H_6$ & 6 & 11 & 9 & 7 & 5 & 3 & 1 & $d=2$\\ \hline
\end{tabular}
\label{H-7}
\end{table}
\end{center}

To design a sequence set with required minimum distance $d_{\req}$ between any two consecutive elements in each of its sequences, we can simply drop all sequences of consecutive minimal distance less than the required distance $d_{\req}$. Alternatively, we may want to drop sequences which contain ``bad" frequencies. This dropping process will be easy if sequence sets are big.

We should remark that ``bad" frequencies are only bad for certain pairs (locations), hence a given sequence which contains a bad frequency for a specific user may not be bad for another user. Then the sequence assignment coordinator can swap sequences between users; if this is not possible, then the sequence can be dropped. This will have a minor effect on the set, especially if the original set of sequences was large.

The example presented in Tables~\ref{S-7} and~\ref{H-7} is suitable for clarifying the construction, however we need a larger example to demonstrate the idea of dropping ``bad" sequences. Prime sequence set for $p = 19$ is shown in Table~\ref{S-19}, and the corresponding HMC sequence set, with dispersed elements, is shown in Table~\ref{H-19}.

\begin{center} 
\begin{table}[!ht]
\caption{The set of prime sequences $S_i$, $1 \leq i \leq 18$ for $p = 19$}
\small
\begin{tabular}{|l||l|l|l|l|l|l|l|l|l|l|l|l|l|l|l|l|l|l|l|} \hline
$S_1 = $ & 0 & 1 & 2 & 3 & 4 & 5 & 6 & 7 & 8 & 9 & 10 & 11 & 12 & 13 & 14 & 15 & 16 & 17 & 18 \\ \hline
$S_2 = $ & 0 & 2 & 4 & 6 & 8 & 10 & 12 & 14 & 16 & 18 & 1 & 3 & 5 & 7 & 9 & 11 & 13 & 15 & 17 \\ \hline
$S_3 = $ & 0 & 3 & 6 & 9 & 12 & 15 & 18 & 2 & 5 & 8 & 11 & 14 & 17 & 1 & 4 & 7 & 10 & 13 & 16 \\ \hline
$S_4 = $ & 0 & 4 & 8 & 12 & 16 & 1 & 5 & 9 & 13 & 17 & 2 & 6 & 10 & 14 & 18 & 3 & 7 & 11 & 15 \\ \hline
$S_5 = $ & 0 & 5 & 10 & 15 & 1 & 6 & 11 & 16 & 2 & 7 & 12 & 17 & 3 & 8 & 13 & 18 & 4 & 9 & 14 \\ \hline
$S_6 = $ & 0 & 6 & 12 & 18 & 5 & 11 & 17 & 4 & 10 & 16 & 3 & 9 & 15 & 2 & 8 & 14 & 1 & 7 & 13 \\ \hline
$S_7 = $ & 0 & 7 & 14 & 2 & 9 & 16 & 4 & 11 & 18 & 6 & 13 & 1 & 8 & 15 & 3 & 10 & 17 & 5 & 12 \\ \hline
$S_8 = $ & 0 & 8 & 16 & 5 & 13 & 2 & 10 & 18 & 7 & 15 & 4 & 12 & 1 & 9 & 17 & 6 & 14 & 3 & 11 \\ \hline
$S_9 = $ & 0 & 9 & 18 & 8 & 17 & 7 & 16 & 6 & 15 & 5 & 14 & 4 & 13 & 3 & 12 & 2 & 11 & 1 & 10 \\ \hline
$S_{10} = $ & 0 & 10 & 1 & 11 & 2 & 12 & 3 & 13 & 4 & 14 & 5 & 15 & 6 & 16 & 7 & 17 & 8 & 18 & 9 \\ \hline
$S_{11} = $ & 0 & 11 & 3 & 14 & 6 & 17 & 9 & 1 & 12 & 4 & 15 & 7 & 18 & 10 & 2 & 13 & 5 & 16 & 8 \\ \hline
$S_{12} = $ & 0 & 12 & 5 & 17 & 10 & 3 & 15 & 8 & 1 & 13 & 6 & 18 & 11 & 4 & 16 & 9 & 2 & 14 & 7 \\ \hline
$S_{13} = $ & 0 & 13 & 7 & 1 & 14 & 8 & 2 & 15 & 9 & 3 & 16 & 10 & 4 & 17 & 11 & 5 & 18 & 12 & 6 \\ \hline
$S_{14} = $ & 0 & 14 & 9 & 4 & 18 & 13 & 8 & 3 & 17 & 12 & 7 & 2 & 16 & 11 & 6 & 1 & 15 & 10 & 5 \\ \hline
$S_{15} = $ & 0 & 15 & 11 & 7 & 3 & 18 & 14 & 10 & 6 & 2 & 17 & 13 & 9 & 5 & 1 & 16 & 12 & 8 & 4 \\ \hline
$S_{16} = $ & 0 & 16 & 13 & 10 & 7 & 4 & 1 & 17 & 14 & 11 & 8 & 5 & 2 & 18 & 15 & 12 & 9 & 6 & 3 \\ \hline
$S_{17} = $ & 0 & 17 & 15 & 13 & 11 & 9 & 7 & 5 & 3 & 1 & 18 & 16 & 14 & 12 & 10 & 8 & 6 & 4 & 2 \\ \hline
$S_{18} = $ & 0 & 18 & 17 & 16 & 15 & 14 & 13 & 12 & 11 & 10 & 9 & 8 & 7 & 6 & 5 & 4 & 3 & 2 & 1 \\ \hline
\end{tabular}
\small
\label{S-19}
\end{table}
\end{center}

\begin{center} 
\begin{table}[!ht]
\caption{The set of HMC sequences $H_i$, $1 \leq i \leq 18$ for $p = 19$ with minimum distance $d$ between consecutive elements}
\tiny
\begin{tabular}{|l||l|l|l|l|l|l|l|l|l|l|l|l|l|l|l|l|l|l|l||l|} \hline
$H_1 = $ & 1 & 3 & 5 & 7 & 9 & 11 & 13 & 15 & 17 & 19 & 21 & 23 & 25 & 27 & 29 & 31 & 33 & 35 & 18 & $d=2$ \\ \hline
$H_2 = $ & 2 & 6 & 10 & 14 & 18 & 22 & 26 & 30 & 34 & 19 & 4 & 8 & 12 & 16 & 20 & 24 & 28 & 32 & 17 & $d = 4$ \\ \hline
$H_3 = $ & 3 & 9 & 15 & 21 & 27 & 33 & 20 & 7 & 13 & 19 & 25 & 31 & 18 & 5 & 11 & 17 & 23 & 29 & 16 & $d =  6$ \\ \hline
$H_4 = $ & 4 & 12 & 20 & 28 & 17 & 6 & 14 & 22 & 30 & 19 & 8 & 16 & 24 & 32 & 21 & 10 & 18 & 26 & 15 & $d =  8$ \\ \hline
$H_5 = $ & 5 & 15 & 25 & 16 & 7 & 17 & 27 & 18 & 9 & 19 & 29 & 20 & 11 & 21 & 31 & 22 & 13 & 23 & 14 & $d =  9$ \\ \hline
$H_6 = $ & 6 & 18 & 30 & 23 & 16 & 28 & 21 & 14 & 26 & 19 & 12 & 24 & 17 & 10 & 22 & 15 & 8 & 20 & 13 & $d =  7$ \\ \hline
$H_7 = $ & 7 & 21 & 16 & 11 & 25 & 20 & 15 & 29 & 24 & 19 & 14 & 9 & 23 & 18 & 13 & 27 & 22 & 17 & 12 & $d =  5$ \\ \hline
$H_8 = $ & 8 & 24 & 21 & 18 & 15 & 12 & 28 & 25 & 22 & 19 & 16 & 13 & 10 & 26 & 23 & 20 & 17 & 14 & 11 & $d =  3$ \\ \hline
$H_9 = $ & 9 & 27 & 26 & 25 & 24 & 23 & 22 & 21 & 20 & 19 & 18 & 17 & 16 & 15 & 14 & 13 & 12 & 11 & 10 & $d =  1$ \\ \hline
$H_{10} = $ & 10 & 11 & 12 & 13 & 14 & 15 & 16 & 17 & 18 & 19 & 20 & 21 & 22 & 23 & 24 & 25 & 26 & 27 & 9 & $d = 1$ \\ \hline
$H_{11} = $ & 11 & 14 & 17 & 20 & 23 & 26 & 10 & 13 & 16 & 19 & 22 & 25 & 28 & 12 & 15 & 18 & 21 & 24 & 8 & $d = 3$ \\ \hline
$H_{12} = $ & 12 & 17 & 22 & 27 & 13 & 18 & 23 & 9 & 14 & 19 & 24 & 29 & 15 & 20 & 25 & 11 & 16 & 21 & 7 & $d = 5$ \\ \hline
$H_{13} = $ & 13 & 20 & 8 & 15 & 22 & 10 & 17 & 24 & 12 & 19 & 26 & 14 & 21 & 28 & 16 & 21 & 30 & 18 & 6 & $d = 7$ \\ \hline
$H_{14} = $ & 14 & 23 & 13 & 22 & 31 & 21 & 11 & 20 & 29 & 19 & 9 & 18 & 27 & 17 & 7 & 16 & 25 & 15 & 5 & $d = 9$ \\ \hline
$H_{15} = $ & 15 & 26 & 18 & 10 & 21 & 32 & 24 & 16 & 8 & 19 & 30 & 22 & 14 & 5 & 17 & 28 & 20 & 12 & 4 & $d = 8$ \\ \hline
$H_{16} = $ & 16 & 29 & 23 & 17 & 11 & 5 & 18 & 31 & 25 & 19 & 13 & 7 & 20 & 33 & 27 & 21 & 15 & 9 & 3 & $d = 6$ \\ \hline
$H_{17} = $ & 17 & 32 & 28 & 24 & 20 & 16 & 12 & 8 & 4 & 19 & 34 & 30 & 26 & 22 & 18 & 14 & 10 & 6 & 2 & $d = 4$ \\ \hline
$H_{18} = $ & 18 & 35 & 33 & 31 & 29 & 27 & 25 & 23 & 21 & 19 & 17 & 15 & 13 & 11 & 9 & 7 & 5 & 3 & 1 & $d = 2$ \\ \hline
\end{tabular}
\tiny
\label{H-19}
\end{table}
\end{center}

Say, now we wanted to design a sequence set with required minimum distance $d_{\req} = 3$. Examining the minimum consecutive distance of each of the 18 one-coincidence sequences of Table~\ref{H-19}, we find four sequences of minimum consecutive distance less than~3: $H_1$, $H_9$, $H_{10}$, and $H_{18}$. Dropping these four sequences, we obtain a set of fourteen sequences each of length 19 and of adjacent distance at least 3. Table~\ref{H-19-1} presents this set.

\begin{center} 
\begin{table}[!ht]
\caption{Set of 14 HMC sequences of period 19 and adjacent distance $ \geq 3$}
\tiny
\begin{tabular}{|l||l|l|l|l|l|l|l|l|l|l|l|l|l|l|l|l|l|l|l||l|} \hline
$H_2 = $ & 2 & 6 & 10 & 14 & 18 & 22 & 26 & 30 & 34 & 19 & 4 & 8 & 12 & 16 & 20 & 24 & 28 & 32 & 17 & $d = 4$ \\ \hline
$H_3 = $ & 3 & 9 & 15 & 21 & 27 & 33 & 20 & 7 & 13 & 19 & 25 & 31 & 18 & 5 & 11 & 17 & 23 & 29 & 16 & $d =  6$ \\ \hline
$H_4 = $ & 4 & 12 & 20 & 28 & 17 & 6 & 14 & 22 & 30 & 19 & 8 & 16 & 24 & 32 & 21 & 10 & 18 & 26 & 15 & $d =  8$ \\ \hline
$H_5 = $ & 5 & 15 & 25 & 16 & 7 & 17 & 27 & 18 & 9 & 19 & 29 & 20 & 11 & 21 & 31 & 22 & 13 & 23 & 14 & $d =  9$ \\ \hline
$H_6 = $ & 6 & 18 & 30 & 23 & 16 & 28 & 21 & 14 & 26 & 19 & 12 & 24 & 17 & 10 & 22 & 15 & 8 & 20 & 13 & $d =  7$ \\ \hline
$H_7 = $ & 7 & 21 & 16 & 11 & 25 & 20 & 15 & 29 & 24 & 19 & 14 & 9 & 23 & 18 & 13 & 27 & 22 & 17 & 12 & $d =  5$ \\ \hline
$H_8 = $ & 8 & 24 & 21 & 18 & 15 & 12 & 28 & 25 & 22 & 19 & 16 & 13 & 10 & 26 & 23 & 20 & 17 & 14 & 11 & $d =  3$ \\ \hline
$H_{11} = $ & 11 & 14 & 17 & 20 & 23 & 26 & 10 & 13 & 16 & 19 & 22 & 25 & 28 & 12 & 15 & 18 & 21 & 24 & 8 & $d = 3$ \\ \hline
$H_{12} = $ & 12 & 17 & 22 & 27 & 13 & 18 & 23 & 9 & 14 & 19 & 24 & 29 & 15 & 20 & 25 & 11 & 16 & 21 & 7 & $d = 5$ \\ \hline
$H_{13} = $ & 13 & 20 & 8 & 15 & 22 & 10 & 17 & 24 & 12 & 19 & 26 & 14 & 21 & 28 & 16 & 21 & 30 & 18 & 6 & $d = 7$ \\ \hline
$H_{14} = $ & 14 & 23 & 13 & 22 & 31 & 21 & 11 & 20 & 29 & 19 & 9 & 18 & 27 & 17 & 7 & 16 & 25 & 15 & 5 & $d = 9$ \\ \hline
$H_{15} = $ & 15 & 26 & 18 & 10 & 21 & 32 & 24 & 16 & 8 & 19 & 30 & 22 & 14 & 5 & 17 & 28 & 20 & 12 & 4 & $d = 8$ \\ \hline
$H_{16} = $ & 16 & 29 & 23 & 17 & 11 & 5 & 18 & 31 & 25 & 19 & 13 & 7 & 20 & 33 & 27 & 21 & 15 & 9 & 3 & $d = 6$ \\ \hline
$H_{17} = $ & 17 & 32 & 28 & 24 & 20 & 16 & 12 & 8 & 4 & 19 & 34 & 30 & 26 & 22 & 18 & 14 & 10 & 6 & 2 & $d = 4$ \\ \hline
\end{tabular}
\tiny
\label{H-19-1}
\end{table}
\end{center}

\bigskip

\section{Properties of HMC sequences}
\label{lemmas}

We now discuss some interesting properties of the sequences $H_k$ that we constructed above over a field $\ff_p$ for a fixed prime $p$. It is clear that the length of each such sequence is $p$ and there are at most $p-1$ sequences in the set. Notice also that, since each prime sequence $S_k$ is uniquely indexed by $k \in \ff_p$, the set of these sequences can be identified with the additive group $\zed/p\zed$. The basic properties of our operations that we will use throughout are, for any integers $a,b,c,d,k$,
\begin{equation}
\label{add_1}
(a+b) \cdot_p k = a \cdot_p k +_p b \cdot_p k \equiv a \cdot_p k + b \cdot_p k\ (\md p),
\end{equation}
and
\begin{equation}
\label{add_2}
(a + b) \cdot_p k + (c + d) \cdot_p k = (a \cdot_p k +_p b \cdot_p k) + (c \cdot_p k +_p d \cdot_p k).
\end{equation}

\begin{lem} \label{H-rep} The elements of a sequence $H_k$ are all distinct.
\end{lem}

\proof
Suppose not, then for some integers $0 \leq i \neq j \leq p-1$ we have 
$$i \cdot_p k + (i+_p1) \cdot_p k = j \cdot_p k + (j+_p1) \cdot_p k.$$
Then the same equality must also hold for addition $\md p$, and so we have an equation in the field $\ff_p$:
$$(2 \cdot_p i +_p 1) \cdot_p k = (2 \cdot_p j+_p1) \cdot_p k,$$
meaning that $2i+1 = 2j+1$ modulo $p$. This means that, viewed as an integer, $(2i+1)-(2j+1)$ is divisible by $p$, i.e. $p \mid 2(i-j)$. Since $p > 2$, this means $p \mid i-j$, however $0 < i-j \leq p-1$, which is a contradiction.
\endproof

\begin{lem} \label{H-size} Each sequence $H_k$ consists of $p$ integers from the set $\{ 1, \dots, 2p-3 \}$.
\end{lem}

\proof
Let $a$ be an element of $H_k$ for some $k \geq 1$, then $a = i \cdot_p k + (i+_p 1) \cdot_p k$ for some $0 \leq i \leq p-1$, $1 \leq k \leq p-1$, where again multiplication is $\md p$ and addition is not. Since $i \cdot_p k, (i +_p 1) \cdot_p k$ are distinct elements of $\ff_p$, 
$$a \leq (p-1) + (p-2) = 2p-3.$$
The conclusion now follows, since elements of $H_k$ are distinct and there are $p$ of them.
\endproof

\begin{lem} \label{H_sum} Let $t \in \zed$ be such that $p=2t+1$. If we write 
$$H_k = \{ a_1,a_2,\dots,a_{t+1},\dots,a_{2t},a_p \},$$
then $a_{t+1} = p$, $a_p = p-a_1$ and for all $2 \leq i < t$, $a_{p-(i-1)} = 2p-a_i$.
\end{lem}

\proof
We number elements of $S_k$ from 0 to $p-1$ and elements of $H_k$ from 1 to~$p$. First notice that for any $1 \leq j \leq p$,
$$j \cdot_p k + (p-j) \cdot_p k = p,$$
i.e. the $j$-th and $(p-j)$-th elements of the prime sequence $S_k$ add up to $p$, while the $0$-the element of $S_k$ is $0$. Then notice that for any $2 \leq i < t$,
\begin{eqnarray*}
a_i + a_{p-(i-1)} & = & ((i-1) \cdot_p k + i \cdot_p k) + ((p - i) \cdot_p k + (p-i +1) \cdot_p k) \\
& = &  ((i-1) \cdot_p k + (p - (i -1)) \cdot_p k) + (i \cdot_p k + (p - i) \cdot_p k) = \\
& = & p + p = 2p.
\end{eqnarray*}
In particular,
$$2p = a_{t+1} + a_{p-(t+1-1)} = a_{t+1} + a_{2t+1-t} = 2a_{t+1},$$
meaning that $a_{t+1} = p$. On the other hand, by~\eqref{add_1}
$$a_p = (p-1) \cdot_p k = p \cdot_p k -_p k = p-k = p - a_1.$$
\endproof

\begin{lem} \label{H_cor} Let $l \neq k$, then for any $0 \leq \tau < p$, $\H_{H_k,H_l}(\tau) \leq 1$.
\end{lem}

\proof
Suppose that for some $0 \leq \tau < p$,
\begin{equation}
\label{H-cor-1}
i \cdot_p k + (i +_p 1) \cdot_p k = (i +_p \tau) \cdot_p l + (i +_p \tau +_p 1) \cdot_p l.
\end{equation}
Reducing modulo $p$, we obtain an equation in the field $\ff_p$:
$$i \cdot_p k +_p (i +_p 1) \cdot_p k = (i +_p \tau) \cdot_p l +_p (i +_p \tau +_p 1) \cdot_p l,$$
which can then be written as
\begin{equation}
\label{H-cor-2}
2 \cdot_p i \cdot_p (k -_p l) = 2 \cdot_p \tau \cdot_p l +_p (l -_p k).
\end{equation}
Equation \eqref{H-cor-2} has precisely one solution in $i$, namely
$$i = \tau \cdot_p l \cdot_p (k -_p l)^{-1} -_p 2^{-1},$$
which means that \eqref{H-cor-1} has at most one solution, and so $\H_{H_k,H_l}(\tau) \leq 1$ for any $0 \leq \tau < p$.
\endproof

\begin{lem} \label{columns} Let $1 \leq k \leq p-1$, and write
$$H_k = \{ a_1,\dots,a_p \},\ H_{p-k} = \{ b_1,\dots,b_p \}.$$
Then for each $1 \leq i \leq p$, $a_i = b_{p-i+1}$. This implies that these two sequences are time-reverse of each other. Further, if we write $H_1,\dots,H_{p-1}$ as rows of a $(p-1) \times p$ table, then the $(p-i)$-th column of this table is the reverse of the $i$-th column.
\end{lem}

\proof
Applying equations~\eqref{add_1} and~\eqref{add_2}, we see that
\begin{eqnarray*}
b_{p-i+1}  & = & (p-i) \cdot_p (p-k) + (p-i+1) \cdot_p (p-k) \\
& = & (p \cdot_p p -_p p \cdot_p k -_p i \cdot_p p +_p i \cdot_p k) \\
& & +\ (p \cdot_p p -_p p \cdot_p k -_p i \cdot_p p +_p i \cdot_p k +_p p -_p k) \\
& = & i \cdot_p k + (i - 1) \cdot_p k = a_i.
\end{eqnarray*}
The implications readily follow.
\endproof

\begin{lem} \label{dist} Let us write $d(H_k)$ for the minimum distance between consecutive elements of $H_k$, and let $p=2t+1$, as above. Then
\begin{enumerate}
\item $d(H_1) = d(H_{p-1}) = 2$,
\item $d(H_t) = d(H_{t+1}) = 1$.
\end{enumerate}
\end{lem} 

\proof
The fact that $d(H_1) = d(H_{p-1})$ and $d(H_t) = d(H_{t+1})$ follows from Lemma~\ref{columns}. Now notice that for $1 \leq j \leq p-1$, $j$-th element of $H_1$ is of the form $a_j = (j-1) + j = 2j-1$ and $a_p = p-1$. Hence any two consecutive elements of $H_1$ are at least two apart, and $a_1,a_2$ are precisely two apart, i.e. $d(H_1) = 2$.

To prove that $d(H_t)=1$, we show that the distance between $a_t$ and $a_{t+1}$ is~1. Notice that
\begin{align*}
a_t & = (t-1) \cdot_p t + t \cdot_p t,\\
a_{t+1} & = (t+1) \cdot_p t + t \cdot_p t.
\end{align*}
Now observe that
$$(t+1) \cdot_p t -_p (t-1) \cdot_p t = \left( (t+1) -_p (t-1) \right) \cdot_p t = 2 \cdot_p t = -1,$$
since $p = 2t+1$. This means that
$$p \mid (t+1) \cdot_p t - (t-1) \cdot_p t + 1,$$
but $\left| (t+1) \cdot_p t - (t-1) \cdot_p t + 1 \right| \leq (p-1) - 1 + 1 < p$, hence we must have 
$$(t+1) \cdot_p t - (t-1) \cdot_p t + 1 = 0.$$
This implies that $a_{t+1} = a_t - 1$, and so we are done.
\endproof
\bigskip

\section{Conclusion}
\label{close}

In this note, we presented a new family of one-coincidence sequence sets with dispersed elements, one for each prime number. These sets are derived from one-coincidence prime sequence sets, such that for each one-coincidence prime sequence set there is a new one-coincidence set comprised of sequences with dispersed sequence elements, required in some circumstances, for FH-CDMA systems~\cite{2}. Each sequence belonging to any constructed set has controlled minimum distance between its adjacent elements. The out-of-phase Hamming auto-correlation function of each sequence belonging to the set is zero. The Hamming cross-correlation between any pair of sequences belonging to any sequence set is at most one. Illustrative examples were presented. Properties of the presented one-coincidence sequence sets are listed and proved. 

The approach adopted in this note can be considered as parallel to the approach of~\cite{4}: both aim to solve the problem of the small distance between adjacent elements, where~\cite{4} does it by making the sequence length smaller than the size of the sequence elements alphabet set, while we reduce sequence set size. Our approach ensures existence of a one-coincidence sequence set for each prime number, while the approach of~\cite{4} does not guarantee the existence of what they call ``generator sequence" for each prime number. 

As a question for future research, it would be interesting to understand the relation between the required minimum consecutive distance $d_{\req}$ of a one-coincidence sequence set, length of sequences, and the sequence set size. For instance, in our example in Table~\ref{H-19} we started with a set of 18 sequences of length 19, and in order to have $d_{\req}=3$ we had to drop 4 sequences, hence obtaining a set of 14 sequences in Table~\ref{H-19-1}. Thus one can ask how big can a one-coincidence set of sequences of prescribed length and $d_{\req}$ be?
\bigskip

{\bf Acknowledgments:} First author was partially supported by the NSA grant H98230-1510051 and by the Simons Foundation grant \#519058. Second author's research at Harvey Mudd College was co-funded by IIE, Mr. Mahesh Kotecha, and Mrs. Mitzi Perdue: for their support he is very grateful. He also acknowledges useful discussions with Prof. R. Olson and Prof. A. Aswad.


\begin{thebibliography}{1}

\bibitem{4}
L.~Bin.
\newblock One-coincidence sequences with specified distance between adjacent
  symbols for frequency-hopping multiple access.
\newblock {\em IEEE Transactions on Communications}, 45(4):408--410, 1997.

\bibitem{2}
S.~Geirhofera, J.~Z. Sunb, L.~Tonga, and B.~M. Sadlerc.
\newblock Cognitive frequency hopping based on interference prediction: Theory
  and experimental results.
\newblock {\em Mobile Computing and Communications Review}, 13(2), 2009.

\bibitem{5}
C.~Hodgdon.
\newblock Adaptive frequency hopping for reduced interference between bluetooth
  and wireless {LAN}.
\newblock {\em Ericsson Technology Licensing}, 2003.

\bibitem{3}
A.~Lempel and H.~Greenberger.
\newblock Families of sequences with optimal {H}amming correlation properties.
\newblock {\em IEEE Transactions on Information Theory}, 20(1):90--94, 1974.

\bibitem{1}
A.~A. Shaar and P.~A. Davies.
\newblock A survey of one-coincidence sequences for frequency-hopped
  spread-spectrum systems.
\newblock {\em IEE Proceedings F - Communications, Radar and Signal
  Processing}, 131(7):719--724, 1984.

\end{thebibliography}
\end{document}